\documentstyle[aps]{revtex}
\def \D {\mbox{D}}
\def \curl {\mbox{curl}\,}
\def \ep {\varepsilon}
\def\be{\begin{equation}}
\def\ee{\end{equation}}
\def\bea{\begin{eqnarray}}
\def\eea{\end{eqnarray}}

\begin{document}

\title{GEOMETRY AND DYNAMICS OF THE BRANE-WORLD\thanks{
Based on an invited talk at EREs2000, Spanish Relativity
Meeting.}}
\author{Roy Maartens\thanks{roy.maartens@port.ac.uk}}
\address{Relativity and Cosmology Group, Portsmouth University,
Portsmouth~PO1~2EG, Britain}

\maketitle

\begin{abstract}

Recent developments in string theory have led to 5-dimensional
warped spacetime models in which standard-model fields are
confined to a 3-brane (the observed universe), while gravity can
propagate in the fifth dimension. Gravity is localized near the
brane at low energies, even if the extra dimension is noncompact.
A review is given of the classical geometry and dynamics of these
brane-world models. The field equations on the brane modify the
general relativity equations in two ways: local 5-D effects are
imprinted on the brane as a result of its embedding, and are
significant at high energies; nonlocal effects arise from the 5-D
Weyl tensor. The Weyl tensor transmits tidal (Coulomb),
gravitomagnetic and gravitational wave effects to the brane from
the 5-D nonlocal gravitational field. Local high-energy effects
modify the dynamics of inflation, and increase the amplitude of
scalar and tensor perturbations generated by inflation. Nonlocal
effects introduce new features in cosmological perturbations. They
induce a non-adiabatic mode in scalar perturbations and massive
modes in vector and tensor perturbations, and they can support
vector perturbations even in the absence of matter vorticity. In
astrophysics, local and nonlocal effects introduce fundamental
changes to gravitational collapse and black hole solutions.

\end{abstract}

\section{Introduction}

At high enough energies, Einstein's theory of general relativity
breaks down and is likely to be a limit of a more general theory.
In string theory/ M theory, gravity is a truly higher-dimensional
theory, becoming effectively 4-dimensional at lower energies.
Recent developments may offer a promising road towards a quantum
gravity theory~\cite{gen}. In brane-world models inspired by
string/M theory, the standard-model fields are confined to a
3-brane, while the gravitational field can propagate in $3+d$
dimensions (the `bulk'). The $d$ extra dimensions need not all be
small, or even compact: recently Randall and Sundrum~\cite{rs}
have shown that for $d=1$, gravity can be localized on a single
3-brane even when the fifth dimension is infinite. This noncompact
localization arises via the exponential `warp' factor in the
non-factorizable metric:
 \be\label{rs}
d\widetilde{s}\,^2=\exp(-2|y|/\ell)\left[-dt^2+d\vec{x}\,^2\right]
+dy^2\,.
 \ee
For $y\neq0$, this metric satisfies the 5-dimensional Einstein
equations with negative 5-dimensional cosmological constant,
$\widetilde{\Lambda}\propto -\ell^{-2}$. The brane is located at
$y=0$, and the induced metric on the brane is a Minkowski metric.
The bulk is a 5-dimensional anti-de Sitter metric, with $y=0$ as
boundary, so that $y<0$ is identified with $y>0$, reflecting the
$Z_2$ symmetry, with the brane as fixed point, that arises in
string theory.

Perturbation of the metric~(\ref{rs}) shows that the Newtonian
gravitational potential on the brane is recovered at lowest order:
 \be\label{newt}
V(r) = {GM\over r}\left(1+{2\ell^2\over 3r^2}\right)+\cdots
 \ee
Thus 4-dimensional gravity is recovered at low energies, with a
first-order correction that is constrained by current
sub-millimetre experiments~\cite{sub}. The lowest order term
corresponds to the massless graviton mode, bound to the brane,
while the corrections arise from massive Kaluza-Klein modes in the
bulk. Generalizing the Randall-Sundrum model to allow for matter
on the brane leads to a generalization of the metric~(\ref{rs}),
and to a breaking of conformal flatness, since matter on the brane
in general induces Weyl curvature in the bulk. Indeed, the massive
Kaluza-Klein modes that produce the corrective terms in
Eq.~(\ref{newt}) reflect the bulk Weyl curvature that arises from
a matter source on the brane.

At a classical level, the brane-world models have a rich
geometrical structure, in which the bulk Weyl curvature tensor and
its interaction with the curvature of the brane play an important
role. The classical geometry and dynamics of brane-world models
are important because they provide a foundation for testing the
astrophysical and cosmological predictions of these models against
observations. Any modification of Einstein's theory that arises in
the quest for quantum gravity needs to pass the tests provided by
increasingly accurate observations. Observers are necessarily
bound to the brane, unable to access the bulk. Bulk effects are
felt indirectly via their imprint on the brane geometry and
dynamics. A covariant Lagrangian approach to the
brane-world~\cite{m} brings out clearly the role of the Weyl
tensors, the kinematical and dynamical quantities, and their
relation to observations.

\section{Field equations on the brane}

Instead of starting from a metric ansatz in special coordinates,
one can generalize the Randall-Sundrum model by a covariant
geometric approach, given by Shiromizu, Maeda and
Sasaki~\cite{sms}. (See also~\cite{class} for related geometrical
results). The unit normal to the brane $n^A$ defines the induced
metric on the brane (and on all hypersurfaces orthogonal to
$n^A$),
 \be
g_{AB}=\widetilde{g}_{AB}-n_An_B\,,
 \ee
where we use tildes to denote the 5-dimensional generalization of
standard general relativity quantities. Without loss of
generality, we can take $n^A$ to be geodesic. A natural choice of
coordinates is $x^A=(x^\mu,y)$, where $x^\mu=(t,x^i)$ are
spacetime coordinates on the brane and $n_A=\delta_A{}^y$. The
extrinsic curvature orthogonal to $n^A$ is
 \be
K_{AB}={\textstyle{1\over2}}{\cal L}_{n}g_{AB}=g_A{}^C
\widetilde{\nabla}_Cn_B\,,
 \ee
so that $K_{[AB]}=0=K_{AB}n^B$, where square brackets denote
anti-symmetrization and ${\cal L}$ is the Lie derivative.

The Gauss equation gives the 4-dimensional curvature tensor as
 \be\label{gauss}
R_{ABCD}= \widetilde{R}_{EFGH}\,g_A{}^Eg_B{}^Fg_C{}^Gg_D{}^H
+2K_{A[C}K_{D]B}\,,
 \ee
and the Codazzi (or Mainardi-Codazzi) equation determines the
change of $K_{AB}$:
 \be\label{cod}
 \nabla_BK^B{}_A-\nabla_AK_B{}^B=\widetilde{R}_{BC}\,g_A{}^Bn^C\,.
 \ee
The 5-dimensional Einstein equations are
\begin{equation}
\widetilde{G}_{AB} =
\widetilde{\kappa}^2\left[-\widetilde{\Lambda}\widetilde{g}_{AB}
+\delta(y)\left\{ -\lambda g_{AB}+T_{AB}\right\}\right]\,,
\label{1}
\end{equation}
where $\widetilde{\kappa}^2= 8\pi/\widetilde{M}_{\rm p}^3$, with
$\widetilde{M}_{\rm p}$ the fundamental 5-dimensional Planck mass,
which is typically much less than the effective Planck mass on the
brane, $M_{\rm p}=1.2\times 10^{19}$ GeV. The brane tension is
$\lambda$, and fields confined to the brane make up the brane
energy-momentum tensor $T_{AB}$, with $T_{AB}n^B=0$. The
Randall-Sundrum model may be further generalized by allowing for a
scalar field in the bulk~\cite{mw}.

Using Eqs.~(\ref{gauss}) and (\ref{1}), it follows that~\cite{sms}
 \bea
{G}_{AB}&=&-{\textstyle{1\over2}}\widetilde{\kappa}^2
\widetilde{\Lambda}g_{AB}+K_C{}^C
K_{AB}-K_A{}^CK_{CB}\nonumber\\&&~{}
+{\textstyle{1\over2}}\left[K^{CD}K_{CD}-\left(K_C{}^C\right)^2
\right]g_{AB} - {\cal E}_{AB}\,,\label{ein}
 \eea
where
 \be
 {\cal E}_{AB} = \widetilde{C}_{ACBD}\,n^Cn^D\,,
 \ee
is the projection of the bulk Weyl tensor orthogonal to $n^A$,
with ${\cal E}_{[AB]}=0={\cal E}_{A}{}^A$. Evaluating
Eq.~(\ref{ein}) on the brane (strictly, as $y\to\pm 0$) will give
the field equations. First, we need to determine $K_{AB}$ at the
brane. The junction conditions across the brane imply that
$g_{AB}$ is continuous, while $K_{AB}$ undergoes a jump due to the
energy-momentum on the brane:
 \be
K_{AB}^{+}-K_{AB}^{-}=-\widetilde{\kappa}^2\left[T_{AB}+
{\textstyle{1\over3}} \left(\lambda-T_C{}^C\right)g_{AB}\right]\,.
 \ee
The $Z_2$ symmetry implies that
 \be\label{z2}
K_{AB}^{-}=-K_{AB}^{+}\,,
 \ee
and then
 \be\label{ext}
K_{AB}=-{\textstyle{1\over2}}
\widetilde{\kappa}^2\left[T_{AB}+{\textstyle{1\over3}}
\left(\lambda-T_C{}^C\right)g_{AB}\right]\,,
 \ee
where we have dropped the $(+)$ and we evaluate quantities on the
brane by taking the limit $y\to+0$.

Finally we arrive at the induced field equations on the
brane~\cite{sms}:
\begin{equation}
G_{\mu\nu}=-\Lambda g_{\mu\nu}+\kappa^2
T_{\mu\nu}+\widetilde{\kappa}^4S_{\mu\nu} - {\cal E}_{\mu\nu}\,,
\label{2}
\end{equation}
where $\kappa^2=8\pi/M_{\rm p}^2$. The energy scales are related
to each other via
\begin{equation}
\lambda=6{\kappa^2\over\widetilde\kappa^4} \,, ~~ \Lambda
={\textstyle{1\over2}}\widetilde\kappa^2
\left(\widetilde{\Lambda}+{\textstyle{1\over6}}\widetilde\kappa^2
\lambda^2\right)\,. \label{3}
\end{equation}
The higher-dimensional modifications of the standard Einstein
equations on the brane are of two forms: first, the matter fields
contribute local quadratic energy-momentum corrections via the
tensor $S_{\mu\nu}$, which arise from the extrinsic curvature, and
second, there are nonlocal effects from the free gravitational
field in the bulk, transmitted via the projection ${\cal
E}_{\mu\nu}$ of the bulk Weyl tensor. The local corrections are
given by
\begin{equation}
S_{\mu\nu}={\textstyle{1\over12}}T_\alpha{}^\alpha T_{\mu\nu}
-{\textstyle{1\over4}}T_{\mu\alpha}T^\alpha{}_\nu+
{\textstyle{1\over24}}g_{\mu\nu} \left[3 T_{\alpha\beta}
T^{\alpha\beta}-\left(T_\alpha{}^\alpha\right)^2 \right]\,.
\label{3'}
\end{equation}
The Weyl tensor $\widetilde{C}_{ABCD}$ represents the free,
nonlocal gravitational field in the bulk. The local part of the
bulk gravitational field is the Einstein tensor
$\widetilde{G}_{AB}$, which is determined locally via the bulk field
equations (\ref{1}). Thus ${\cal E}_{\mu\nu}$ {\em transmits
nonlocal gravitational degrees of freedom from the bulk to the
brane, including tidal (or Coulomb), gravito-magnetic and
transverse traceless (gravitational wave) effects.}

There may be other branes in the bulk. Branes interact
gravitationally via any Weyl curvature that they generate. On the
observer's brane at $y=0$, the presence of other branes is felt
indirectly through their contribution to ${\cal E}_{\mu\nu}$.

As a consequence of the Codazzi equation~(\ref{cod}), the form of
the bulk energy-momentum tensor (which means that
$\widetilde{R}_{BC}\,g_A{}^Bn^C=0$) and $Z_2$ symmetry, it follows
that the brane energy-momentum tensor is conserved:
\begin{equation}\label{5'}
\nabla^\nu T_{\mu\nu}=0 \,.
\end{equation}
When there are scalar and other fields in the bulk, this is no
longer in general true~\cite{mw,bdbl}, and non-conservation of
$T_{\mu\nu}$ reflects an exchange of energy-momentum between the
brane and the bulk. In the case when there is only a cosmological
constant in the bulk, there is no such exchange. Using
Eq.~(\ref{5'}) in Eq.~(\ref{2}), the contracted Bianchi identities
on the brane, $\nabla^\mu G_{\mu\nu}=0$, imply that the projected
Weyl tensor obeys the constraint
\begin{equation}
\nabla^\mu{\cal E}_{\mu\nu}={6\kappa^2\over\lambda}\,\nabla^\mu
S_{\mu\nu}\,. \label{5}
\end{equation}
This shows that ${\cal E}_{\mu\nu}$ is sourced by energy-momentum
terms, which in general include spatial gradients and time
derivatives. Thus evolution and inhomogeneity in the matter fields
can generate nonlocal gravitational effects in the bulk, which
then `backreact' on the brane.

The dynamical equations on the brane are equations~(\ref{2}),
(\ref{5'}) and (\ref{5}). It is important to note that these
equations are not in general closed on the brane~\cite{sms}, since
Eq.~(\ref{5}) does not determine ${\cal E}_{\mu\nu}$ in general,
as further discussed below.  This reflects the fact that there are
bulk degrees of freedom which cannot be predicted from data
available on the brane, for example, incoming gravitational
radiation which impinges on the brane. One needs to solve the
field equations in the bulk in order to fully determine ${\cal
E}_{\mu\nu}$ on the brane.

\section{Covariant interpretation of gravity localization}

In the Randall-Sundrum model, with a flat brane, localization of
gravity at the brane is understood perturbatively via
Eq.~(\ref{newt}). For a general matter distribution on a curved
brane, we can provide a qualitative non-perturbative and covariant
interpretation of gravity localization via tidal acceleration.

Consider a field of observers on the brane (e.g., observers
comoving with matter) with 4-velocity $u^\mu$, and let $u^A$ be an
extension off the brane (the result does not depend on the
extension), so that $u^An_A=0$, $u^Au_A=-1$. The tidal
acceleration in the $n^A$ direction measured by the observers is
$-n_A\widetilde{R}^{\,A}{}_{BCD}{u}^Bn^C {u}^D$. Now
 \be\label{rc} \widetilde{R}_{ABCD}=
\widetilde{C}_{ABCD}+{\textstyle{2\over3}}
\left\{\widetilde{g}_{A[C}\widetilde{R}_{D]B}+
\widetilde{g}_{B[D}\widetilde{R}_{C]A}\right\}-{\textstyle{1\over6}}
\widetilde{R}\widetilde{g}_{A[C}\widetilde{g}_{D]B}\,,
 \ee
so that by the field equation~(\ref{1}) (and recalling that
$T_{AB}n^B=0$),
 \be\label{rc2}
-\widetilde{R}_{ABCD}n^A {u}^Bn^C {u}^D=-{\cal E}_{AB}{u}^A{u}^B+
{\textstyle{1\over6}}\widetilde{\kappa}^2\widetilde{\Lambda}\,.
 \ee
Taking the limit $y\to+0$, we get~\cite{m}
 \be
\mbox{tidal acceleration in off-brane direction }=
{\textstyle{1\over6}}\widetilde{\kappa}^2 \widetilde{\Lambda}
-{\cal E}_{\mu\nu}u^\mu u^\nu \,. \label{6b}
 \ee
Since $\widetilde{\Lambda}<0$, it contributes to acceleration {\em
towards} the brane. This reflects the confining role of the
negative bulk cosmological constant on the gravitational field in
the generalized Randall-Sundrum type models. Equation~(\ref{6b})
also shows that localization of the gravitational field near the
brane is enhanced if ${\cal E}_{\mu\nu}u^\mu u^\nu>0$, which
corresponds to a negative effective energy density on the brane
from nonlocal bulk effects.

\section{Covariant decomposition of local and nonlocal bulk effects}

The general form of the brane energy-momentum tensor for any
matter fields (scalar fields, perfect fluids, kinetic gases,
dissipative fluids, etc.), including a combination of different
fields, can be covariantly given as
\begin{equation}
T_{\mu\nu}=\rho u_\mu
u_\nu+ph_{\mu\nu}+\pi_{\mu\nu}+q_{\mu}u_{\nu}+q_\nu u_\mu\,.
 \label{3''}
\end{equation}
Here $\rho$ and $p$ are the energy density and isotropic pressure,
and $h_{\mu\nu}=g_{\mu\nu}+u_\mu u_\nu$ projects orthogonal to
$u^\mu$ on the brane. The energy flux obeys
$q_{\mu}=q_{\langle\mu\rangle}$, and the anisotropic stress obeys
$\pi_{\mu\nu}=\pi_{\langle\mu\nu\rangle}$, where angled brackets
denote the projected, symmetric and tracefree part:
 \be
V_{\langle\mu\rangle}=h_\mu{}^\nu V_\nu\,,~~
W_{\langle\mu\nu\rangle}=\left[h_{(\mu}{}^\alpha h_{\nu)}{}^\beta-
{\textstyle{1\over3}}h^{\alpha\beta}h_{\mu\nu}\right]W_{\alpha\beta}\,,
 \ee
with round brackets denoting symmetrization. In an inertial frame
at any point on the brane, we have $u^\mu=\delta^\mu{}_0$ and
$h_{\mu\nu}={\rm diag}(0,1,1,1)$, $q_\mu=(0,q_i)$, $\pi_{\mu0}=0$.

The tensor $S_{\mu\nu}$, which carries local bulk effects onto the
brane, may then be irreducibly decomposed as
\begin{eqnarray}
&&S_{\mu\nu}={\textstyle{1\over24}}\left[2\rho^2-3\pi_{\alpha\beta}
\pi^{\alpha\beta}\right]u_\mu u_\nu
+{\textstyle{1\over24}}\left[2\rho^2+4\rho p+\pi_{\alpha\beta}
\pi^{\alpha\beta}-4q_\alpha q^\alpha\right]h_{\mu\nu} \nonumber\\
&&{}~~-
{\textstyle{1\over12}}(\rho+2p)\pi_{\mu\nu}+\pi_{\alpha\langle\mu}
\pi_{\nu\rangle}{}^\alpha +q_{\langle\mu}q_{\nu\rangle}+
{\textstyle{1\over3}}\rho q_{(\mu}u_{\nu)}- {\textstyle{1\over12}}
q^\alpha \pi_{\alpha(\mu}u_{\nu)} \,. \label{3'''}
\end{eqnarray}
This simplifies for a perfect fluid or minimally-coupled scalar
field:
 \be
S_{\mu\nu}={\textstyle{1\over12}}\rho^2 u_\mu u_\nu
+{\textstyle{1\over12}}\rho\left(\rho+2 p\right)h_{\mu\nu}\,.
 \ee
The quadratic energy-momentum corrections to standard general
relativity will thus be significant for
$\widetilde{\kappa}^4\rho^2
> 12\kappa^2\rho$, i.e., in the high-energy regime
 \be
\rho > \lambda \sim \left({\widetilde{M}_{\rm p}\over M_{\rm
p}}\right)^2 \widetilde{M}_{\rm p}^{\,4}\,.
 \ee
The lower bound arising from current tests for deviations from
Newton's law is~\cite{mwbh}
 \be
\widetilde{M}_{\rm p}> 10^5~{\rm TeV}\,,~~\lambda^{1/4} > 100~{\rm
GeV}\,.
 \ee
(A much weaker limit is imposed by nucleosynthesis constraints.)

Nonlocal effects from the bulk are encoded in the brane tensor
${\cal E}_{\mu\nu}$, which can be decomposed as
\begin{equation}
{\cal E}_{\mu\nu}=-{6\over\kappa^2\lambda}\left[{\cal
U}\left(u_\mu u_\nu+{\textstyle {1\over3}} h_{\mu\nu}\right)+{\cal
P}_{\mu\nu}+{\cal Q}_{\mu}u_{\nu}+{\cal Q}_\nu u_\mu\right]\,.
\label{6}
\end{equation}
The factor $6/\kappa^2\lambda$, which is
$(\widetilde\kappa/\kappa)^4$ by Eq.~(\ref{3}), is introduced for
dimensional reasons, and also since it ensures that in the general
relativity limit, $\lambda^{-1}\to0$, we have ${\cal
E}_{\mu\nu}\to0$. We have written ${\cal E}_{\mu\nu}$ as an
effective energy-momentum tensor: the bulk Weyl tensor imprints on
the brane an effective energy density, stresses and energy flux.

The effective nonlocal energy density on the brane, arising from
the free gravitational field in the bulk, is
\[
{\cal U}=-({\textstyle{1\over6}}\kappa^2\lambda)\, {\cal
E}_{\mu\nu}u^\mu u^\nu\,.
\]
This nonlocal energy density need not be
positive~\cite{ssm,dmpr,d}. Since ${\cal E}_{\mu\nu}$ is
tracefree, the effective nonlocal pressure is ${1\over3}{\cal U}$.
There is an effective nonlocal anisotropic stress
\[
{\cal P}_{\mu\nu}= -({\textstyle{1\over6}}\kappa^2\lambda)\, {\cal
E}_{\langle\mu\nu\rangle}
\]
on the brane, arising from the free gravitational field in the
bulk, and
\[
{\cal Q}_\mu =-({\textstyle{1\over6}}\kappa^2\lambda)\,  {\cal
E}_{\langle\mu\rangle\nu}u^\nu
\]
is an effective nonlocal energy flux on the brane, arising from
the free gravitational field in the bulk.

If the bulk is anti-de Sitter (AdS$_5$), as in the Randall-Sundrum
model, then ${\cal E}_{\mu\nu}=0$, since the bulk is conformally
flat:
 \be
\mbox{AdS$_5$ bulk:}~~{\cal E}_{\mu\nu}=0\,.
 \ee
The Randall-Sundrum model has a Minkowski brane, but AdS$_5$ can
also admit a Friedmann brane, and satisfy the Einstein
equations~(\ref{1}). However, the most general solution with a
Friedmann brane is Schwarzschild-anti de Sitter
spacetime~\cite{msm}. Then it follows from the Friedmann
symmetries that
 \be
\mbox{SAdS$_5$ bulk, Friedmann brane:}~~{\cal Q}_\mu=0={\cal
P}_{\mu\nu}\,,
 \ee
where ${\cal U}=0$ only if the mass of the black hole in the bulk
is zero. The presence of the black hole leads to a `dark
radiation' term in the Friedmann equation (see below). For a
static spherically symmetric brane (e.g. a static stellar interior
or the exterior of a nonrotating black hole)~\cite{dmpr}:
 \be
\mbox{static spherical brane:}~~{\cal Q}_\mu=0\,.
 \ee
This condition also holds for a Bianchi~I brane~\cite{mss}.

The local and nonlocal bulk corrections may be consolidated into
an effective total energy density, pressure, anisotropic stress
and energy flux, since the modified field equations~(\ref{2}) take
the standard Einstein form with a redefined energy-momentum
tensor:
\begin{equation}
G_{\mu\nu}=-\Lambda g_{\mu\nu}+\kappa^2 T^{\rm tot}_{\mu\nu}\,,
\label{6'}
\end{equation}
where
\begin{equation}
T^{\rm tot}_{\mu\nu}= T_{\mu\nu}+{6\over \lambda}S_{\mu\nu}-
{1\over\kappa^2}{\cal E}_{\mu\nu}\,. \label{6''}
\end{equation}
Then it follows from Eqs.~(\ref{3'''}) and (\ref{6}) that
\begin{eqnarray}
\rho^{\rm tot} &=& \rho+{1\over 4\lambda}\left(2\rho^2 -3
\pi_{\mu\nu}\pi^{\mu\nu}\right) +{6\over \kappa^4\lambda}{\cal U}
\label{a}\\ p^{\rm tot} &=& p+ {1\over
4\lambda}\left(2\rho^2+4\rho p+ \pi_{\mu\nu}\pi^{\mu\nu}-4q_\mu
q^\mu\right) +{2\over \kappa^4\lambda}{\cal U} \label{b}\\
\pi^{\rm tot}_{\mu\nu} &=& \pi_{\mu\nu}+{1\over 2\lambda}
\left[-(\rho+3p)\pi_{\mu\nu}+
\pi_{\alpha\langle\mu}\pi_{\nu\rangle}{}^\alpha+q_{\langle\mu}q_
{\nu\rangle}\right] + {6\over \kappa^4\lambda}{\cal
P}_{\mu\nu}\label{c}\\ q^{\rm tot}_\mu &=&q_\mu+{1\over 4\lambda}
\left(4\rho q_\mu-\pi_{\mu\nu}q^\nu\right)+ {6\over
\kappa^4\lambda}{\cal Q}_\mu \,.\label{d}
\end{eqnarray}
These general expressions make clear the local and nonlocal
effects. They simplify in the case of a perfect fluid (or
minimally coupled scalar field, or isotropic one-particle
distribution function), i.e., for $q_\mu=0=\pi_{\mu\nu}$. However,
we note that the total energy flux and anisotropic stress do not
vanish in this case in general:
 \be
q^{\rm tot}_\mu ={6\over \kappa^4\lambda}{\cal Q}_\mu\,,~~
\pi^{\rm tot}_{\mu\nu} ={6\over \kappa^4\lambda}{\cal
P}_{\mu\nu}\,.
 \ee
Nonlocal bulk effects can contribute to effective imperfect fluid
terms even when the matter on the brane has perfect fluid form.

\section{Local and nonlocal conservation equations}

The brane energy-momentum tensor and the consolidated effective
energy-momentum tensor are {\em both} conserved separately, by
virtue of Eqs.~(\ref{5'}) and (\ref{5}). Conservation of
$T_{\mu\nu}$ gives the standard general relativity energy and
momentum conservation equations
\begin{eqnarray}
&&\dot{\rho}+\Theta(\rho+p)+\D^\mu q_\mu+2A^\mu
q_\mu+\sigma^{\mu\nu}\pi_{\mu \nu}=0\,,\label{c1}\\ &&
\dot{q}_{\langle\mu\rangle}+{\textstyle{4\over3}}\Theta
q_\mu+\D_\mu p+(\rho+p)A_\mu + \D^\nu
\pi_{\mu\nu}+A^\nu\pi_{\mu\nu} \nonumber\\&&~~~~{}
+\sigma_{\mu\nu}q^\nu- [\omega, q]_\mu=0\,.\label{c2}
\end{eqnarray}
In these equations, an overdot denotes $u^\nu\nabla_\nu$,
$\Theta=\nabla^\mu u_\mu$ is the volume expansion rate of the
$u^\mu$ congruence, $A_\mu=\dot{u}_\mu =A_{\langle\mu\rangle}$ is
its 4-acceleration,
$\sigma_{\mu\nu}=\D_{\langle\mu}u_{\nu\rangle}$ is its shear rate,
and $\omega_\mu=-{1\over2}\curl u_\mu=\omega_{\langle\mu\rangle}$
is its vorticity rate. On a Friedmann brane,
$A_\mu=\omega_\mu=\sigma_{\mu\nu}=0$ and $\Theta=3H$, where
$H=\dot a/a$ is the Hubble rate.

The covariant spatial curl is given by~\cite{m1}
 \be
\curl V_\mu=\ep_{\mu\alpha\beta}\D^\alpha V^\beta\,,~~ \curl W
_{\mu\nu}=\ep_{\alpha\beta(\mu}\D^\alpha W^\beta{}_{\nu)}\,,
 \ee
where $\ep_{\mu\nu\sigma}$ is the projection orthogonal to $u^\mu$
of the brane alternating tensor, and $\D_\mu$ is the projected
part of the brane covariant derivative, defined by
 \be
\D_\mu F^{\alpha\cdots}{}{}_{\cdots\beta}=\left(\nabla_\mu
F^{\alpha\cdots}{}{}_{\cdots\beta}\right)_\perp= h_\mu{}^\nu
h^\alpha{}_\gamma \cdots h_\beta{}^\delta \nabla_\nu
F^{\gamma\cdots}{}{}_{\cdots\delta}\,.
 \ee
The covariant cross-product is
$[V,Y]_\mu=\ep_{\mu\alpha\beta}V^\alpha Y^\beta$. In a local
inertial frame at a point on the brane, with
$u^\mu=\delta^\mu{}_0$, we have: $0=A_0=\omega_0=\sigma_{0\mu}=
\ep_{0\mu\nu}=[V,Y]_0= \curl V_0 =\curl W_{0\mu}$ and $\D_\mu
F^{\alpha\cdots}{}{}_{\cdots\beta}=\delta_\mu{}^i
\delta^\alpha{}_j\cdots\delta_\beta{}^k\nabla_i
F^{j\cdots}{}{}_{\cdots k}$.

The conservation of $T^{\rm tot}_{\mu\nu}$ gives, upon using Eqs.
(\ref{a})--(\ref{c2}), what we can call nonlocal conservation
equations~\cite{m}. The nonlocal energy conservation equation is a
propagation equation for ${\cal U}$:
\begin{eqnarray}
&& \dot{\cal U}+{\textstyle{4\over3}}\Theta{\cal U}+\D^\mu{\cal
Q}_\mu+2A^\mu{\cal Q}_\mu+\sigma^{\mu\nu}{\cal
P}_{\mu\nu}\nonumber\\ &&~{}={\textstyle{1\over24}}\kappa^4\left[
6\pi^{\mu\nu}\dot{\pi}_{\mu\nu}+6(\rho+p)\sigma^{\mu\nu}
\pi_{\mu\nu}+2\Theta \left(2q^\mu q_\mu+
\pi^{\mu\nu}\pi_{\mu\nu}\right) +2A^\mu
q^\nu\pi_{\mu\nu}\right.\nonumber\\ &&~\left.{}
-4q^\mu\D_\mu\rho+q^\mu\D^\nu\pi_{\mu\nu} +\pi^{\mu\nu}\D_\mu
q_\nu -2\sigma^{\mu\nu}\pi_{\alpha\mu}\pi_\nu{}^\alpha-
2\sigma^{\mu\nu}q_\mu q_\nu \right]\,. \label{c1'}
\end{eqnarray}
The nonlocal momentum conservation equation is a propagation
equation for ${\cal Q}_\mu$:
\begin{eqnarray}
&& \dot{\cal
Q}_{\langle\mu\rangle}+{\textstyle{4\over3}}\Theta{\cal Q}_\mu
+{\textstyle{1\over3}}\D_\mu{\cal U}+{\textstyle{4\over3}}{\cal
U}A_\mu +\D^\nu{\cal P}_{\mu\nu}+A^\nu{\cal
P}_{\mu\nu}+\sigma_{\mu\nu}{\cal Q}^\nu-[\omega,{\cal Q}]_\mu
\nonumber\\ &&~~{}={\textstyle{1\over24}} \kappa^4 \left[
-4(\rho+p)\D_\mu \rho +6(\rho+p)\D^\nu\pi_{\mu\nu}
+q^\nu\dot{\pi}_{\langle\mu\nu\rangle}+\pi_\mu{}^\nu
\D_\nu(2\rho+5p)\right.\nonumber\\
&&~~~\left.{}-{\textstyle{2\over3}}\pi^{\alpha\beta}
\left(\D_\mu\pi_{\alpha
\beta}+3\D_\alpha\pi_{\beta\mu}\right)-3\pi_{\mu\alpha}\D_\beta
\pi^{\alpha\beta}+{\textstyle{28\over3}}q^\nu\D_\mu q_\nu
\right.\nonumber\\ &&~~~\left.{}+4\rho
A^\nu\pi_{\mu\nu}-3\pi_{\mu\alpha}A_\beta\pi^{\alpha\beta}
+{\textstyle{8\over3}}A_\mu\pi^{\alpha\beta}\pi_{\alpha\beta}
-\pi_{\mu\alpha}\sigma^{\alpha\beta}q_\beta\right.\nonumber\\
&&~~~\left.{}+\sigma_{\mu\alpha} \pi^{\alpha\beta}q_\beta+
\pi_{\mu\nu}[\omega, q]^\nu -\ep_{\mu\alpha\beta}\omega^\alpha
\pi^{\beta\nu}q_\nu+4(\rho+p)\Theta
q_\mu\right.\nonumber\\&&~~~\left.{}+ 6q_\mu A^\nu q_\nu
+{\textstyle{14\over3}}A_\mu q^\nu q_\nu+4q_\nu
\sigma^{\alpha\beta} \pi_{\alpha\beta}\right]\,.\label{c2'}
\end{eqnarray}
All of the matter source terms on the right of these two
equations, except for the first term on the right of
Eq.~(\ref{c2'}), are imperfect fluid terms, and most of these
terms are quadratic in the imperfect quantities $q_\mu$ and
$\pi_{\mu\nu}$. For perfect fluid matter, only the $\D_\mu\rho$
term on the right of Eq.~(\ref{c2'}) survives, but in realistic
cosmological and astrophysical models, further terms will survive.
For example, terms linear in $\pi_{\mu\nu}$ will carry the photon
quadrupole in cosmology or the shear viscous stress in stellar
models.

In general, the 4 independent equations determine 4 of the 9
independent components of ${\cal E}_{\mu\nu}$ on the brane. What
is missing, is an evolution equation for ${\cal P}_{\mu\nu}$
(which has up to 5 independent components). Thus in general, the
projection of the 5-dimensional field equations onto the brane,
together with $Z_2$ matching, does not lead to a closed system.
Nor could we expect this to be the case, since there are bulk
degrees of freedom whose impact on the brane cannot be predicted
by brane observers. Our decomposition of ${\cal E}_{\mu\nu}$ has
shown that {\em the evolution of the nonlocal energy density and
flux (carrying scalar and vector modes from bulk gravitons) is
determined on the brane, while the evolution of the nonlocal
anisotropic stress (carrying tensor, as well as scalar and vector,
modes) is not.}

In special cases the missing equation does not matter. For
example, if ${\cal P}_{\mu\nu}=0$, as in the case of a Friedmann
brane, then the evolution of ${\cal E}_{\mu\nu}$ is determined by
Eqs. (\ref{c1'}) and (\ref{c2'}). If the brane is stationary (with
Killing vector parallel to $u^\mu$), then evolution equations are
not needed for ${\cal E}_{\mu\nu}$. However, small perturbations
of these special cases will immediately restore the problem of
missing information.

If the matter on the brane has a perfect-fluid energy-momentum
tensor, the local conservation equations~(\ref{c1}) and (\ref{c2})
reduce to
\begin{eqnarray}
&&\dot{\rho}+\Theta(\rho+p)=0\,,\label{pc1}\\ && \D_\mu
p+(\rho+p)A_\mu =0\,,\label{pc2}
\end{eqnarray}
while the nonlocal conservation equations~(\ref{c1'}) and
(\ref{c2'}) reduce to
\begin{eqnarray}
&& \dot{\cal U}+{\textstyle{4\over3}}\Theta{\cal U}+\D^\mu{\cal
Q}_\mu+2A^\mu{\cal Q}_\mu+\sigma^{\mu\nu}{\cal P}_{\mu\nu}=0\,,
\label{pc1'}\\&& \dot{\cal
Q}_{\langle\mu\rangle}+{\textstyle{4\over3}}\Theta{\cal Q}_\mu
+{\textstyle{1\over3}}\D_\mu{\cal U}+{\textstyle{4\over3}}{\cal
U}A_\mu +\D^\nu{\cal P}_{\mu\nu}+A^\nu{\cal
P}_{\mu\nu}+\sigma_{\mu\nu}{\cal Q}^\nu-[\omega,{\cal Q}]_\mu
\nonumber\\&&~~{} =-{\textstyle{1\over6}} \kappa^4 (\rho+p)\D_\mu
\rho\,.\label{pc2'}
\end{eqnarray}

Equation (\ref{pc2'}) shows that~\cite{sms} {\em if ${\cal
E}_{\mu\nu}=0$ and the brane energy-momentum tensor has perfect
fluid form, then the density $\rho$ must be homogeneous}. The
converse does not hold, i.e., homogeneous density does {\em not}
in general imply vanishing ${\cal E}_{\mu\nu}$. A simple example
is the Friedmann case: Eq. (\ref{pc2'}) is trivially satisfied,
while Eq. (\ref{pc1'}) becomes
 \be
\dot{\cal U}+4H{\cal U}=0\,.
 \ee
This equation has the `dark radiation' solution
\begin{equation}\label{dr}
{\cal U}={\cal U}_o\left({a_o\over a}\right)^4\,.
\end{equation}
If ${\cal E}_{\mu\nu}=0$, then the field equations on the brane
form a closed system. Thus {\em for perfect fluid branes with
homogeneous density and ${\cal E}_{\mu\nu}=0$, the brane field
equations form a consistent closed system.} However, there is no
guarantee that the resulting brane metric can be embedded in a
regular bulk.

It also follows as a corollary that {\em inhomogeneous density
requires nonzero} ${\cal E}_{\mu\nu}$. For example, stellar
solutions on the brane necessarily have ${\cal E}_{\mu\nu}\neq0$
in the stellar interior if it is non-uniform. Perturbed Friedmann
models on the brane also must have ${\cal E}_{\mu\nu}\neq0$. Thus
a nonzero ${\cal E}_{\mu\nu}$ is inevitable in realistic
astrophysical and cosmological models.

For a perfect fluid at very high energies, i.e., $\rho\gg\lambda$,
and for which we can neglect ${\cal U}$ (e.g., in an inflating
cosmology), Eqs.~(\ref{a}) and (\ref{b}) show that~\cite{gm}
 \bea
w^{\rm tot} &\equiv & {p^{\rm tot}\over\rho^{\rm tot}}\approx 2w+1
\,,\label{vh1}\\ (c_{\rm s}^2)^{\rm tot}&\equiv & {\dot{p}^{\rm
tot}\over\dot{\rho}^{\rm tot}}\approx c_{\rm s}^2
+w+1\,,\label{vh2}
 \eea
where $w=p/\rho$ and $c_{\rm s}^2=\dot p/\dot\rho$. Thus {\em at
very high energies on the brane, the effective equation of state
and sound speed are stiffened.} This can have important
consequences in the early universe and during gravitational
collapse. For example, in a very high-energy radiation era,
$w={1\over3}$, the effective cosmological equation of state is
ultra-stiff: $w^{\rm tot}\approx {5\over3}$. In late-stage
gravitational collapse of pressureless matter, $w=0$, the
effective equation of state is stiff, $w^{\rm tot}\approx 1$, and
the effective pressure is nonzero and dynamically important.

\section{Gravitational collapse to a black hole on the brane}

The dynamics of gravitational collapse on the brane is not yet
properly understood, given the complications that are introduced
by local high-energy effects and nonlocal effects. In general
relativity, it is known that nonrotating collapse to a black hole
leads uniquely to the Schwarzschild metric, with no `hair', i.e.,
no trace of the dynamics of the collapse process. On the brane,
this is no longer true. {\em The Schwarzschild metric cannot
describe the final state of collapse}~\cite{chr,gkr,dmpr}, since
it does not incorporate the 5-dimensional behaviour of the
gravitational potential in the strong-field regime (the metric is
incompatible with massive Kaluza-Klein modes). The appropriate
metric on the brane to describe the black hole final state is not
known, but a non-perturbative exterior solution should have {\em
nonzero} ${\cal E}_{\mu\nu}$ in order to be compatible with
massive Kaluza-Klein modes in the strong-field regime. In the
end-state of collapse, we expect a nonlocal field ${\cal
E}_{\mu\nu}$ which goes to zero at large distances, recovering the
Schwarzschild weak-field limit, but which grows at short range.
Furthermore, ${\cal E}_{\mu\nu}$ may carry a Weyl `fossil' record
of the collapse process.

The known black hole solutions on the brane include the
Schwarzschild solution, for which ${\cal E}_{\mu\nu}=0$, but also
other solutions with nonzero nonlocal effects. A vacuum on the
brane satisfies the field equations
 \be\label{vac}
R_{\mu\nu}=-{\cal E}_{\mu\nu}\,,~~ R^\mu{}_\mu=0={\cal
E}^\mu{}_\mu \,,~~ \nabla^\nu{\cal E}_{\mu\nu}=0\,.
 \ee
It follows that {\em Einstein-Maxwell solutions in general
relativity produce vacuum solutions on the brane, where the
electromagnetic field is replaced by a nonlocal Weyl
energy-momentum tensor}~\cite{dmpr}. Examples are the
Reissner-N\"ordstrom-like solution discussed below, and a
Vaidya-like solution~\cite{dg}. The equations~(\ref{vac}) form a
closed system on the brane in the stationary case, including the
static spherical case, for which
 \be
\Theta=0=\omega_\mu=\sigma_{\mu\nu}\,,~~\dot{\cal U}=0={\cal
Q}_\mu=\dot{\cal P}_{\mu\nu} \,.
 \ee
The nonlocal conservation equations reduce to
 \be
{\textstyle{1\over3}}{\D}_{\mu}{\cal U}+{\textstyle{4\over3}}{\cal
U}A_\mu + \D^\nu{\cal P}_{\mu\nu}+A^\nu{\cal P}_{\mu\nu}=0\,,
 \ee
and the general solution of this and the remaining brane field
equations, with metric of the form,
 \be
ds^2=-F(r)dt^2+F^{-1}(r)dr^2+r^2d\Omega^2\,,\label{bh2}
 \ee
is then~\cite{dmpr}
 \bea
F&=&1-\left({2M\over M_{\rm p}^2}\right){1\over r} +\left(
{q\over\widetilde{M}_{\rm p}^2 }\right){1\over r^2}\,,\label{bh}\\
{\cal E}_{\mu\nu}&=&-\left( {q\over\widetilde{M}_{\rm p}^2
}\right){1\over r^4}\left[u_\mu u_\nu-2r_\mu r_\nu+h_{\mu\nu}
\right]\,,
 \eea
where $r_\mu$ is a unit radial vector.

This solution has the form of the general relativity
Reissner-Nordstr\"om solution, but there is {\em no electric
field} on the brane. Instead, the nonlocal Coulomb effects
imprinted by the bulk Weyl tensor have induced a {\em `tidal'
charge} parameter $q$. This parameter depends on the mass $M$ on
the brane, i.e., $q=q(M)$, since this is the source of bulk Weyl
field (leaving aside the more complicated case where there may be
additional Weyl sources in the bulk). In order to preserve the
spacelike nature of the singularity, we need $q<0$. This is in
accord with the intuitive idea that the tidal charge {\em
strengthens} the gravitational field, since it arises from the
source mass $M$ on the brane. By contrast, in the
Reissner-Nordstr\"om solution of general relativity, $q=+Q^2$
weakens the gravitational field. Negative tidal charge, $q< 0$,
means that there is one horizon on the brane, outside the
Schwarzschild horizon:
 \be
r_{+}={M\over M_{\rm p}^2}\left[1+\sqrt {1 -q{M_{\rm p}^4 \over
M^2 \widetilde{M}_{\rm p}^2}}\,\right]> r_{\rm s}\,.
 \ee

The tidal-charge black hole metric does not satisfy the far-field
$r^{-3}$ correction to the gravitational
potential~\cite{rs,gkr,bper,ssm}, as in Eq.~(\ref{newt}), and
therefore cannot describe the end-state of collapse. However,
Eq.~(\ref{bh}) shows the correct 5-dimensional behaviour of the
potential ($\propto r^{-2}$) at short distances, so that the
tidal-charge metric should be a good approximation in the
strong-field regime. This is the regime where the Schwarzschild
metric on the brane fails, whereas the Schwarzschild metric is a
good approximation at large distances.

The bulk solution corresponding to the tidal-charge brane metric
is not known, although numerical investigations have been
performed~\cite{crss}. For the Schwarzschild special case $q=0$,
the brane is the surface $z=1$ in AdS$_5$, in conformal
coordinates~\cite{chr}:
 \be\label{schw}
d\widetilde{s}\,^2=\left({3\over \kappa^2\lambda}\right){1\over
z^2}\left[ds^2+dz^2\right]\,.
 \ee
Here $ds^2$ is the 4-dimensional Schwarzschild metric, given by
Eqs.~(\ref{bh2}) and (\ref{bh}) with $q=0$. The singularity $r=0$
is a line along the $z$-axis, so that this bulk metric describes a
`black string'. The 5-dimensional horizon is the surface
$g_{tt}=0$ in the bulk, which is $r=2M/M_{\rm p}^2$. This bulk
horizon is a sphere of radius $r=r_{\rm s}$ on each $z=$constant
surface, so that it has a cylindrical shape in the $z$-direction.
The bulk is singular at the AdS$_5$ horizon, and the black string
horizon is unstable~\cite{chr,g}.

If the physically realistic black hole bulk metric has the form
 \be
d\widetilde{s}\,^2= -F(r,y)dt^2+J(r,y)dr^2+L(r,y)
r^2d\Omega^2+dy^2\,,
 \ee
in Gaussian normal coordinates, then the bulk horizon is the
surface $F(r,y)=0$. Perturbative studies~\cite{gkr,g} and exact
lower-dimensional solutions~\cite{ehm1} have been used to probe
the shape of the bulk horizon, but a non-perturbative 5-D solution
has not yet been found.

Clearly the black hole solution, and the collapse process that
leads to it, have a far richer structure in the brane-world than
in general relativity. Another intriguing issue is how Hawking
radiation is modified by bulk effects~\cite{ehm}.

\section{Exact cosmological solutions}

By Eqs.~(\ref{dr}) and (\ref{gc2}), the generalized Friedmann
equation on a spatially homogeneous and isotropic brane
is~\cite{bdel,bcos}
\begin{equation}\label{f}
H^2={\textstyle{1\over3}}\kappa^2\rho\left(1+{\rho\over
2\lambda}\right)+{\textstyle{1\over3}}\Lambda -{K\over a^2} +
{2{\cal U}_o\over\kappa^2\lambda} \left({a_o\over a}\right)^4\,,
\end{equation}
where $K=0,\pm1$. The nonlocal term that arises from bulk Coulomb
effects (also called the dark radiation term) is strongly limited
by nucleosynthesis~\cite{bdel,lmsw}:
 \be
{1\over \kappa^2\lambda}\left({ {\cal U}\over \rho}\right)_{\rm
nucl}< 0.005\,.
 \ee
This means that the nonlocal term is always sub-dominant during
the radiation era, and rapidly becomes negligible after the
radiation era.

If we neglect the nonlocal term, and take $\Lambda=0=K$, then the
Friedmann equation during very high-energy inflation and early
radiation-domination becomes
 \be
H^2 \approx {\kappa^2\over6} \,{\rho^2\over\lambda}\,.
 \ee
The enhanced expansion rate, compared to general relativity (where
$H\propto\sqrt{\rho}$), introduces important changes to the
dynamics of the early universe~\cite{mwbh,inf,cs} and leads to an
{\em increase in the amplitude of scalar}~\cite{mwbh} {\em and
tensor}~\cite{lmw} {\em perturbations generated by inflation.} If
$\rho\gg\lambda$ at the start of the radiation era, then the
solution of the Friedmann equation gives $a\propto t^{1/4}$.

The generalized Raychaudhuri equation~(\ref{pr}) (with
$\Lambda=0={\cal U}$) reduces to
 \be
\dot H+H^2= - {\textstyle{1\over6}}
\kappa^2\left[\rho+3p+(2\rho+3p) {\rho\over \lambda}\right]\,,
 \ee
and shows that the condition for inflation on the brane
($\ddot{a}>0$, i.e. $\dot H+H^2>0$) is~\cite{mwbh}
 \be
w<-{1\over3}\left({2\rho+\lambda\over \rho+\lambda}\right)\,.
 \ee
As $\rho/\lambda\to0$, we recover the general relativity result,
$w<-{1\over3}$. But in the very high-energy regime, we find
$w<-{2\over3}$.

The dynamics of homogeneous but anisotropic Bianchi branes has
also been investigated~\cite{mss,cs}. In particular, high-energy
bulk effects can strongly alter the dynamics near the singularity,
because matter can dominate over shear anisotropy, opposite to the
case of general relativity~\cite{mss}.

The bulk metric for a flat Friedmann brane may be given
explicitly~\cite{bdel}:
 \be\label{fex}
d\widetilde{s}\,^2= -N(t,y)^2dt^2+{A}(t,y)^2d\vec{x}\,^2+dy^2\,,
 \ee
where $A(t,0)=a(t)$ and $t$ is proper time on the brane, so that
$N(t,0)=1$. The metric functions are
\begin{eqnarray}
N&=&{\dot {A}\over\dot{a}}\,,\label{fex1}\\ {A}^2&=&
-{\rho\over2\lambda}\left(2+{\rho\over \lambda}\right)
a^2-{6\,{\cal U}_oa_o^4\over\lambda^2a^2}-\left(1+
{\rho\over\lambda}\right)a^2\,\sinh(2\mu|y|)
\nonumber\\&&~{}+\left[{1\over2}\left(2+{2\rho\over\lambda}+
{\rho^2\over\lambda^2}\right) a^2+{6\,{\cal
U}_oa_o^4\over\lambda^2a^2}\right]\cosh(2\mu y)\,,\label{fex2}
\end{eqnarray}
with
% \be
$\mu=\kappa^3\sqrt{6/\lambda}\,.$
% \ee
This bulk metric is Schwarzschild-AdS$_5$ spacetime~\cite{msm} (in
Gaussian normal coordinates), with the mass parameter of the black
hole in the bulk proportional to ${\cal U}_o$.

\section{Propagation and constraint equations}

Section V gave propagation equations for the local and nonlocal
energy density ${\cal U}$ and flux ${\cal Q}_\mu$. The remaining
covariant equations on the brane are the propagation and
constraint equations for the kinematic quantities $\Theta$,
$A_\mu$, $\omega_\mu$, $\sigma_{\mu\nu}$, and for the nonlocal
gravitational field on the brane. The kinematic quantities govern
the relative motion of neighbouring fundamental world-lines. The
nonlocal gravitational field on the brane is given by the {\em
brane} Weyl tensor $C_{\mu\nu\alpha\beta}$. This splits into the
gravito-electric and gravito-magnetic fields on the brane:
 \be
E_{\mu\nu}=C_{\mu\alpha\nu\beta}u^\alpha u^\beta
=E_{\langle\mu\nu\rangle }\,,~~
H_{\mu\nu}={\textstyle{1\over2}}\ep_{\mu\alpha \beta}
C^{\alpha\beta}{}{}_{\nu\gamma}u^\gamma=H_{\langle\mu\nu\rangle}
\,,
 \ee
where $E_{\mu\nu}$ must not be confused with ${\cal E}_{\mu\nu}$.
The Ricci identity for $u^\mu$ and the Bianchi identities
$\nabla^\beta C_{\mu\nu\alpha\beta} =
\nabla_{[\mu}(-R_{\nu]\alpha} + {1\over6}Rg_{\nu]\alpha})$ produce
the fundamental evolution and constraint equations governing the
above covariant quantities~\cite{ee}. The field equations are
incorporated via the algebraic replacement of the Ricci tensor
$R_{\mu\nu}$ by the effective total energy-momentum tensor,
according to Eq.~(\ref{6'}). The brane equations are derived
directly from the standard general relativity versions~\cite{cov}
by simply replacing the energy-momentum tensor terms $\rho,\dots$
by $\rho^{\rm tot},\dots$. For a perfect fluid or
minimally-coupled scalar field, the general equations~\cite{m}
reduce to the following.\\

\noindent Generalized Raychaudhuri equation (expansion
propagation):
\begin{eqnarray}
&&\dot{\Theta}+{\textstyle{1\over3}}\Theta^2+\sigma_{\mu\nu}
\sigma^{\mu\nu} -2\omega_\mu\omega^\mu -{\rm D}^\mu A_\mu+A_\mu
A^\mu+{\textstyle{1\over2}}\kappa^2(\rho + 3p) -\Lambda
\nonumber\\&&~~{}=
-{\textstyle{1\over2}}\kappa^2(2\rho+3p){\rho\over\lambda}-
{6\over\kappa^2\lambda}{\cal U}\,. \label{pr}
\end{eqnarray}
Vorticity propagation:
\begin{equation}
\dot{\omega}_{\langle \mu\rangle }
+{\textstyle{2\over3}}\Theta\omega_\mu +{\textstyle{1\over2}}\curl
A_\mu -\sigma_{\mu\nu}\omega^\nu=0 \,.\label{pe4}
\end{equation}
Shear propagation:
\begin{equation}
\dot{\sigma}_{\langle \mu\nu \rangle }
+{\textstyle{2\over3}}\Theta\sigma_{\mu\nu}
+E_{\mu\nu}-\D_{\langle \mu}A_{\nu\rangle } +\sigma_{\alpha\langle
\mu}\sigma_{\nu\rangle }{}^\alpha+ \omega_{\langle
\mu}\omega_{\nu\rangle} - A_{\langle \mu}A_{\nu\rangle}
={3\over\kappa^2\lambda}{\cal P}_{\mu\nu}\,. \label{pe5}
\end{equation}
Gravito-electric propagation (Maxwell-Weyl E-dot equation):
\begin{eqnarray}
 && \dot{E}_{\langle \mu\nu \rangle }
+\Theta E_{\mu\nu} -\curl H_{\mu\nu}
+{\textstyle{1\over2}}\kappa^2(\rho+p)\sigma_{\mu\nu}
\nonumber\\&&~{} -2A^\alpha\ep_{\alpha\beta(\mu}H_{\nu)}{}^\beta
-3\sigma_{\alpha\langle \mu}E_{\nu\rangle }{}^\alpha
+\omega^\alpha \ep_{\alpha\beta(\mu}E_{\nu)}{}^\beta
\nonumber\\&&~~{}= -{\textstyle{1\over2}}\kappa^2
(\rho+p){\rho\over\lambda}\sigma_{\mu\nu}
-{1\over\kappa^2\lambda}\left[4{\cal U}\sigma_{\mu\nu}+3\dot{\cal
P}_{\langle \mu\nu \rangle} +\Theta{\cal P}_{\mu\nu}
+3\D_{\langle\mu}{\cal Q}_{\nu\rangle}
\right.\nonumber\\&&~~~\left.{} +6A_{\langle\mu}{\cal
Q}_{\nu\rangle}+ 3\sigma^\alpha{}_{\langle\mu} {\cal
P}_{\nu\rangle\alpha}+3 \omega^\alpha\ep_{\alpha\beta(\mu} {\cal
P}_{\nu)}{}^\beta\right] \,. \label{pe6}
\end{eqnarray}
Gravito-magnetic propagation (Maxwell-Weyl H-dot equation):
\begin{eqnarray}
 &&\dot{H}_{\langle
\mu\nu \rangle } +\Theta H_{\mu\nu} +\curl E_{\mu\nu}-
3\sigma_{\alpha\langle \mu}H_{\nu\rangle }{}^\alpha +\omega^\alpha
\ep_{\alpha\beta(\mu}H_{\nu)}{}^\beta
+2A^\alpha\ep_{\alpha\beta(\mu}E_{\nu)}{}^\beta \nonumber\\&&~~{}=
{3\over\kappa^2\lambda}\left[ \curl{\cal
P}_{\mu\nu}-3\omega_{\langle\mu} {\cal Q}_{\nu\rangle}
+\sigma^\alpha{}_{(\mu}\ep_{\nu)\alpha\beta} {\cal Q}^\beta\right]
\,. \label{pe7}
\end{eqnarray}
Vorticity constraint:
\begin{equation}
\D^\mu\omega_\mu -A^\mu\omega_\mu =0\,.\label{pcc1}
\end{equation}
Shear constraint:
\begin{equation}
\D^\nu\sigma_{\mu\nu}-\curl\omega_\mu
-{\textstyle{2\over3}}\D_\mu\Theta +2[\omega,A]_\mu =
-{6\over\kappa^2\lambda} {\cal Q}_\mu
 \,.\label{pcc2}
\end{equation}
Gravito-magnetic constraint:
\begin{equation}
 \curl\sigma_{\mu\nu}+\D_{\langle \mu}\omega_{\nu\rangle  }
 -H_{\mu\nu}+2A_{\langle \mu}
\omega_{\nu\rangle  }=0 \,.\label{pcc3}
\end{equation}
\newpage\noindent Gravito-electric divergence (Maxwell-Weyl div-E equation) :
\begin{eqnarray}
 && \D^\nu E_{\mu\nu}
 -{\textstyle{1\over3}}\kappa^2\D_\mu\rho
 -[\sigma,H]_\mu
+3H_{\mu\nu}\omega^\nu \nonumber\\&&{}= {\kappa^2 \rho\over
3\lambda} \D_\mu\rho +{1\over\kappa^2\lambda}\left(2\D_\mu{\cal
U}-2\Theta{\cal Q}_\mu-3\D^\nu{\cal P}_{\mu\nu}
+3\sigma_{\mu\nu}{\cal Q}^\nu-9[\omega,{\cal Q}]_\mu\right)\!,
\label{pcc4}
\end{eqnarray}
where the covariant tensor commutator is $[W,Z]_\mu
=\ep_{\mu\alpha\beta}W^\alpha{}_\gamma Z^{\beta\gamma}$.\\

\noindent Gravito-magnetic divergence (Maxwell-Weyl div-H
equation):
\begin{eqnarray}
 &&\D^\nu H_{\mu\nu}
-\kappa^2(\rho+p)\omega_\mu +[\sigma,E]_\mu
 -3E_{\mu\nu}\omega^\nu
\nonumber\\&&~~{}=\kappa^2(\rho+ p){\rho\over\lambda} \omega_\mu +
{1\over\kappa^2\lambda}\left(8 {\cal U} \omega_\mu-3\curl{\cal
Q}_\mu-3[\sigma,{\cal P}]_\mu-3{\cal P}_{\mu\nu}\omega^\nu\right)
\,.\label{pcc5}
\end{eqnarray}
Gauss-Codazzi equations on the brane ($\omega_\mu=0$)~\cite{mss}:
 \bea
&&R^\perp_{\langle \mu\nu\rangle}+\dot{\sigma}_{\langle \mu\nu
\rangle }+\Theta\sigma_{\mu\nu} -\D_{\langle \mu}A_{\nu\rangle }
-A_{\langle \mu}A_{\nu\rangle}={6\over\kappa^2\lambda}{\cal
P}_{\mu\nu}\,, \label{gc1}\\ &&R^\perp+
{\textstyle{2\over3}}\Theta^2-\sigma_{\mu\nu} \sigma^{\mu\nu}
-2\kappa^2\rho -2\Lambda = {\kappa^2\rho^2\over\lambda}+
{12\over\kappa^2\lambda}{\cal U}\,, \label{gc2}
 \eea
where $R^\perp_{\mu\nu}$ is the Ricci tensor for 3-surfaces
orthogonal to $u^\mu$ on the brane and
$R^\perp=h^{\mu\nu}R^\perp_{\mu\nu}$.\\

The standard 4-dimensional general relativity results are regained
when $\lambda^{-1}\to0$, which sets all right hand sides to zero
in Eqs.~(\ref{pr})--(\ref{gc2}). Together with
Eqs.~(\ref{pc1})--(\ref{pc2'}), {\em these equations govern the
dynamics of the matter and gravitational fields on the brane,
incorporating both the local (quadratic energy-momentum) and
nonlocal (projected 5-D Weyl) effects from the bulk.} Local terms
are proportional to $\rho/\lambda$, and are significant only at
high energies. Nonlocal terms contain ${\cal U}$, ${\cal Q}_\mu$
and ${\cal P}_{\mu\nu}$, with the latter two quantities
introducing imperfect fluid effects, even though the matter has
perfect fluid form.

Bulk effects give rise to important new driving and source terms
in the propagation and constraint equations. The vorticity
propagation and constraint, and the gravito-magnetic constraint
have no direct bulk effects, but all other equations do. Local and
nonlocal energy density are driving terms in the expansion
propagation. The spatial gradients of local and nonlocal energy
density provide sources for the gravito-electric field. The
nonlocal anisotropic stress is a driving term in the propagation
of shear and the gravito-electric/ -magnetic fields, and the
nonlocal energy flux is a source for shear and the
gravito-magnetic field. The Maxwell-Weyl equations show in detail
the contribution to the nonlocal gravito-electromagnetic field on
the brane, i.e., $(E_{\mu\nu},H_{\mu\nu})$, from the nonlocal
5-dimensional Weyl field in the bulk.

The system of propagation and constraint equations, i.e.
Eqs.~(\ref{pc1})--(\ref{pc2'}) and (\ref{pr})--(\ref{gc2}), is
exact and nonlinear, applicable to both cosmological and
astrophysical modelling, including strong-gravity effects. In the
next section we will linearize the system in order to study
cosmological perturbations on the brane. A different linearization
scheme could be developed for studying compact objects.

In general the system of equations is not closed: there is no
evolution equation for the nonlocal anisotropic stress ${\cal
P}_{\mu\nu}$. As noted above, closure on the brane can arise in
special cases, when the nonlocal conservation
equations~(\ref{c1'}) and (\ref{c2'}) are sufficient to determine
${\cal E}_{\mu\nu}$. These special cases include:\\

${\cal P}_{\mu\nu}=0$, as on a Friedmann brane;

${\cal E}_{\mu\nu}=0$ and $\D_\mu\rho=0$, with perfect fluid
matter, as in special cases of Friedmann and Bianchi branes;

stationary brane, with $\dot{\cal E}_{\mu\nu}=0$, so that
Eqs.~(\ref{c1'}) and (\ref{c2'}) reduce to constraint equations;

branes with isotropic 3-Ricci curvature,
$\omega_\mu=0=R^\perp_{\langle\mu\nu\rangle}$.\\

The last case follows from Eq.~(\ref{gc1}), which becomes an
equation defining ${\cal P}_{\mu\nu}$ in terms of quantities
already determined via other equations. Special cases are
Friedmann and Bianchi~I branes. In general, for anisotropic
3-Ricci curvature, Eq.~(\ref{gc1}) is an equation determining
$R^\perp_{\langle\mu\nu\rangle}$ in terms of other quantities.

Although the special cases above can give consistent closure on
the brane, there is no guarantee that the brane is embeddable in a
regular bulk. This is the case for a Friedmann brane, whose
symmetries imply (together with $Z_2$ matching) that the bulk is
Schwarzschild-AdS$_5$~\cite{msm}. A Schwarzschild brane can be
embedded in a `black string' bulk metric, but this has
singularities~\cite{chr,g}.

\newpage

\section{Cosmological perturbations}

The dynamics of the background homogeneous brane provides
important constraints on brane-world cosmologies. Roughly
speaking, these constraints amount to the statement that the
brane-world should reproduce the standard Friedmann dynamics from
nucleosynthesis onwards. However, much stricter constraints are
implied by the growing body of data on the cosmic microwave
background (CMB) radiation and large-scale structure (LSS), since
this data indirectly probes earlier times. In order to confront
brane-world models with the data, we need to study perturbations
on the brane.

As is apparent from the exact nonlinear equations in the previous
section, cosmological perturbations on the brane are qualitatively
more complicated than in general relativity. Not only is the
background more complicated, but effects from the bulk that are
imprinted on the brane include degrees of freedom that cannot be
predicted from data available to brane observers. A complete
understanding of brane perturbations therefore necessarily
involves the analysis of bulk perturbations -- for which the mode
equations are {\em partial} differential equations, with
nontrivial boundary conditions.

A covariant brane-perturbation theory has been developed~\cite{m}
and applied to large-scale density perturbations~\cite{gm}.
Metric-based general formalisms for bulk perturbations have been
developed~\cite{new}, and large-scale perturbations generated from
quantum fluctuations during de Sitter inflation on the brane have
also been partly computed~\cite{mwbh,hhr,gs,lmw,bmw}. Large-scale
scalar perturbations and their impact on the CMB have been
analysed~\cite{lmsw}.

The results so far may be summarized as follows:\\

\noindent {\sf Tensor perturbations:}\, {\em Bulk effects produce
a massless mode during inflation and a continuum of massive
Kaluza-Klein modes}~\cite{gs,lmw}, {\em with} $m>{3\over2}H$. The
massive modes stay in the vacuum state, and on large scales there
is a constant mode with enhanced amplitude (compared to general
relativity)~\cite{lmw}. We expect no qualitative change on large
scales to 4-dimensional general relativity tensor modes in the CMB
and LSS, but there could be a significant change on small scales
due to the massive modes.\\

\noindent {\sf Vector perturbations:}\, {\em Bulk effects can
support vector perturbations, even without matter
vorticity}~\cite{m,bmw}, but these modes, which do not arise in
general relativity, are massive (there is no normalizable massless
mode), and stay in the vacuum state during inflation on the
brane~\cite{bmw}. The momentum can be determined on large scales
without solving the bulk perturbations, but {\em the vector
Sachs-Wolfe effect cannot be found on-brane}~\cite{bmw}, because
of the nonlocal anisotropic stress, which is undetermined on the
brane. There are possibly qualitative changes to 4-dimensional
general relativity vector modes in the CMB and LSS on large
scales.
\\

\noindent {\sf Scalar perturbations:} \, {\em Bulk effects
introduce a non-adiabatic mode on large scales}~\cite{m,gm,lmsw}.
Density perturbations on large scales can be solved on-brane,
without solving for the bulk perturbations~\cite{gm}, but {\em the
Sachs-Wolfe effect cannot be found on-brane}~\cite{lmsw}, because
of the nonlocal anisotropic stress, which is undetermined on the
brane. There are possibly qualitative changes to 4-dimensional
general relativity scalar modes in the CMB and LSS, even on large
scales, and probably significant changes on small scales.\\

The quantitative results up to now are confined to large scales.
Further progress requires the integration of the bulk mode
equations. In the simplest case, for tensor perturbations on a
flat Friedmann brane, the mode equation is~\cite{lmw}
 \be
{\partial\over\partial t}\left({A^3\over N}{\partial {\cal H}\over
\partial t}\right)- {\partial\over\partial y} \left( A^3N
{\partial {\cal H}\over
\partial y}\right)+ k^2AN{\cal H}=0\,,
 \ee
where $A$ and $N$ are given by Eqs.~(\ref{fex1}) and (\ref{fex2}).
The perturbed metric, in Gaussian normal coordinates, is
 \be
d\widetilde{s}\,^2= -N(t,y)^2dt^2+{A}(t,y)^2\left[ \delta_{ij}+
{\cal H}_{ij}(t,\vec{x},y)\right] d{x}^idx^j+dy^2\,,
 \ee
where ${\cal H}_i{}^i=0=\partial^j{\cal H}_{ij}$ and
 \be
 {\cal H}_{ij}(t,\vec{x},y)={\cal H}(t,y)\exp({\rm i}
 \vec{k}\cdot \vec{x})\, {\rm e}_{ij}\,,
 \ee
with ${\rm e}_{ij}$ a polarization tensor. At the brane, the
boundary condition is
 \be
\left({\partial {\cal H}\over \partial y}\right)_{y=0}=\pi^{\sc
t}\,,
 \ee
where $\pi^{\sc t}(t)\exp({\rm i} \vec{k}\cdot \vec{x})\, {\rm
e}_{ij}$ is the anisotropic stress of matter on the brane.\\

In order to illustrate some of the features of perturbations on
the brane, we can use the brane-based covariant approach~\cite{m}.
The exact nonlinear equations in previous sections can be
linearized as follows. The limiting case of the background
Friedmann brane is characterized by the vanishing of all
inhomogeneous and anisotropic quantities:
 \be
\D_\mu f=V_\mu=W_{\mu\nu}=0\,,
 \ee
where $f=\rho,p,\Theta,{\cal U}$, and
$V_\mu=A_\mu,\omega_\mu,{\cal Q}_\mu$, and
$W_{\mu\nu}=\sigma_{\mu\nu}, E_{\mu\nu}, H_{\mu\nu}, {\cal
P}_{\mu\nu}$. These quantities are then first-order of smallness
in the linearization scheme, and since they vanish in the
background, they are gauge-invariant~\cite{ehb}. The linearized
conservation equations (assuming adiabatic matter perturbations)
are
\begin{eqnarray}
&&\dot{\rho}+\Theta(\rho+p)=0\,,\label{ppc1}\\ &&  c_{\rm
s}^2\D_\mu \rho+(\rho+p)A_\mu =0\,,\label{ppc2} \\ && \dot{\cal
U}+{\textstyle{4\over3}}\Theta{\cal U}+\D^\mu{\cal Q}_\mu  =0 \,,
\label{lc1'}\\&& \dot{\cal Q}_{\mu}+4H{\cal Q}_\mu
+{\textstyle{1\over3}}\D_\mu{\cal U}+{\textstyle{4\over3}}{\cal
U}A_\mu +\D^\nu{\cal P}_{\mu\nu} =-{\textstyle{1\over6}}
\kappa^4(\rho+p) \D_\mu \rho \,.\label{lc2'}
\end{eqnarray}
Linearization of the propagation and constraint equations leads
to:
\begin{eqnarray}
&&\dot{\Theta}+{\textstyle{1\over3}}\Theta^2 -{\rm D}^\mu
A_\mu+{\textstyle{1\over2}}\kappa^2(\rho + 3p) -\Lambda
=
-{\textstyle{1\over2}}\kappa^2(2\rho+3p){\rho\over\lambda}-
{6\over\kappa^2\lambda}{\cal U}\,, \label{prl}\\ && \dot{\omega}_{
\mu } +2H\omega_\mu +{\textstyle{1\over2}}\curl A_\mu =0
\,,\label{pe4l}\\ && \dot{\sigma}_{ \mu\nu } +2H\sigma_{\mu\nu}
+E_{\mu\nu}-\D_{\langle \mu}A_{\nu\rangle }
={3\over\kappa^2\lambda} {\cal P}_{\mu\nu}\,, \label{pe5l}\\ &&
\dot{E}_{ \mu\nu  } +3H E_{\mu\nu} -\curl H_{\mu\nu}
+{\textstyle{1\over2}}\kappa^2(\rho+p)\sigma_{\mu\nu} =
-{\textstyle{1\over2}}\kappa^2
(\rho+p){\rho\over\lambda}\sigma_{\mu\nu}\nonumber\\&&~~{}
-{1\over\kappa^2\lambda}\left[4{\cal U}\sigma_{\mu\nu}+3\dot{\cal
P}_{\langle \mu\nu \rangle} +\Theta{\cal P}_{\mu\nu}
+3\D_{\langle\mu}{\cal Q}_{\nu\rangle} \right] \,, \label{pe6l}\\
&&\dot{H}_{ \mu\nu
 } +3H H_{\mu\nu} +\curl E_{\mu\nu}
={3\over\kappa^2\lambda} \curl{\cal P}_{\mu\nu} \,, \label{pe7l}
\\&&\D^\mu\omega_\mu =0\,,\label{pcc1l}\\
&&\D^\nu\sigma_{\mu\nu}-\curl\omega_\mu
-{\textstyle{2\over3}}\D_\mu\Theta  = -{6\over\kappa^2\lambda}
{\cal Q}_\mu
 \,,\label{pcc2l}\\ &&
\curl\sigma_{\mu\nu}+\D_{\langle \mu}\omega_{\nu\rangle  }
 -H_{\mu\nu}=0 \,,\label{pcc3l}\\ && \D^\nu E_{\mu\nu}
 -{\textstyle{1\over3}}\kappa^2\D_\mu\rho
={\kappa^2 \rho\over 3\lambda} \D_\mu\rho
+{1\over\kappa^2\lambda}\left[2\D_\mu{\cal U}-2\Theta{\cal
Q}_\mu-3\D^\nu{\cal P}_{\mu\nu}\right] \,,\label{pcc4l}\\
 &&\D^\nu H_{\mu\nu}
-\kappa^2(\rho+p)\omega_\mu = \kappa^2(\rho+ p){\rho\over\lambda}
\omega_\mu + {1\over\kappa^2\lambda}\left[8 {\cal U}
\omega_\mu-3\curl{\cal Q}_\mu\right] \,.\label{pcc5l}
\end{eqnarray}
Equations~(\ref{ppc1}), (\ref{lc1'}) and (\ref{prl}) do not
provide gauge-invariant equations for perturbed quantities, but
their spatial gradients do.

A covariant, gauge-invariant (and purely local) splitting into
scalar, vector and tensor modes is given by (compare~\cite{bert})
 \bea
 V_\mu &=& \D_\mu V+\bar{V}_\mu\,,\label{pp} \\
 W_{\mu\nu} &=& \D_{\langle\mu}\D_{\nu\rangle}{W}
+\D_{\langle\mu}\bar{W}_{\nu\rangle}+\bar{W}_{\mu\nu}\,, \label{p}
 \eea
where an overbar denotes a transverse (divergence-free) quantity
(note that $W_{\mu\nu}$ is already tracefree). Purely scalar modes
are characterized by
 \be
 \bar{V}_\mu=\bar{W}_\mu=\bar{W}_{\mu\nu}=0\,,
 \ee
and standard identities~\cite{m}, the vorticity constraint
equation (\ref{pcc1l}) and the gravito-magnetic constraint
equation (\ref{pcc3l}) then show that
\begin{equation}\label{s1}
\curl V_\mu=0=\curl W_{\mu\nu}\,,~ \D^\nu
W_{\mu\nu}={\textstyle{2\over3}}\D^2( \D_\mu
W)\,,~\omega_\mu=0=H_{\mu\nu}\,.
\end{equation}
\newpage\noindent
Vector modes obey
 \be\label{vec2}
V_\mu=\bar{V}_\mu\,,~~W_{\mu\nu}=\D_{\langle\mu}
\bar{W}_{\nu\rangle}\,,~~\curl\D_\mu f=-2\dot{f}\omega_\mu\,,
 \ee
and it then follows that
 \be\label{vec}
\D^\nu W_{\mu\nu}={\textstyle{1\over2}}\D^2\bar{W}_\mu\,,~~ \curl
W_{\mu\nu}={\textstyle{1\over2}}
\D_{\langle\mu}\bar{W}_{\nu\rangle}\,.
 \ee
Tensor modes are covariantly characterized by
 \be
\D_\mu f=0=V_\mu\,,~~ W_{\mu\nu}=\bar{W}_{\mu\nu}\,.
 \ee

\subsection{Density perturbations on the brane}

We define the density and expansion scalars (as in general
relativity~\cite{ehb})
\begin{equation}\label{s3}
\Delta={a^2\over\rho}\D^2\rho\,,~~Z=a^2\D^2\Theta\,,
\end{equation}
and scalars describing inhomogeneity in the nonlocal quantities:
\begin{equation}\label{s4}
U={a^2\over\rho}\D^2{\cal U}\,,~~Q={a\over\rho} \D^2 {\cal
Q}\,,~~P={1\over \rho}\D^2{\cal P}\,.
\end{equation}
Then we take the comoving spatial Laplacian of Eqs.~(\ref{ppc1}),
(\ref{lc1'}) and (\ref{prl}), using standard identities~\cite{m}.
This leads to a system of 4 evolution equations for $\Delta$, $Z$,
$U$, and $Q$. In general the system is under-determined, since
there is no evolution equation for $P$. However, $P$ only arises
via its Laplacian, and on large scales, the {\em system closes and
brane observers can predict density perturbations from initial
conditions intrinsic to the brane.} The system on large scales is
\begin{eqnarray}
\dot{\Delta} &=&3wH\Delta-(1+w)Z\,,\label{s5}\\ \dot{Z}
&=&-2HZ-{6\rho\over\kappa^2\lambda} U %\nonumber\\ &&~~{}
-{\textstyle{1\over2}}\kappa^2\rho\left[1+
(4+3w){\rho\over\lambda} - {12c_{\rm s}^2{\cal U}\over
(1+w)\kappa^4\lambda\rho}\right] \Delta\,,\label{s6}\\ \dot{U}
&=&(3w-1)HU - \left({4c_{\rm s}^2\over 1+w}\right)\left({{\cal
U}\over\rho}\right) H\Delta -\left({4{\cal U}\over3\rho}\right)
Z\,,\label{s7}\\ \dot{Q} &=&(1-3w)HQ-{1\over3a}U+{1\over6a}\left[
{8c_{\rm s}^2{\cal U}\over (1+w)\rho}-\kappa^4
\rho^2(1+w)\right]\Delta\,.\label{s8}
\end{eqnarray}
In standard general relativity, only the first two equations
apply, with $\lambda^{-1}$ set to zero in Eq.~(\ref{s6}), so that
we can decouple the density perturbations via a second-order
equation for $\Delta$, whose independent solutions are adiabatic
growing and decaying modes. {\em Bulk effects introduce a new
non-adiabatic mode:} there are 3 coupled equations in $\Delta$,
$Z$ and $U$, plus a decoupled equation for $Q$, which is
determined once the other 3 quantities are solved for. Some
solutions are found in~\cite{gm} for ${\cal U}=0$ in the
background, and they show how the fluctuations $U$ in ${\cal U}$
introduce effective entropy perturbations, and how $\Delta$
evolves differently from the standard general relativity case when
$\rho\gg\lambda$ (see also~\cite{lmsw}).

\subsection{Vector perturbations on the brane}

The linearized vorticity propagation equation (\ref{pe4l}) does
not carry any bulk effects, and vorticity decays with expansion as
in standard general relativity, reflecting the fact that angular
momentum conservation holds on the brane. The gravito-magnetic
divergence equation~(\ref{pcc5l}), becomes
 \be
\D^2\bar{H}_{\mu}= 2
\kappa^2(\rho+p)\left[1+{\rho\over\lambda}\right] \omega_\mu +
{2\over\kappa^2\lambda}\left[8 {\cal U} \omega_\mu-3\curl\bar{\cal
Q}_\mu\right]\,,
 \ee
on using Eq.~(\ref{vec}). This shows that the local and nonlocal
energy density and the nonlocal energy flux provide additional
sources for the brane gravito-magnetic field. In standard general
relativity, it is necessary to increase the angular momentum
density $(\rho+p)\omega_\mu$ in order to increase the
gravito-magnetic field, but bulk effects allow additional sources.
In particular, unlike in general relativity, {\em it is possible
to source vector perturbations even when the vorticity vanishes,}
since $\curl\bar{\cal Q}_\mu$ may be nonzero.

We can find a closed system of equations on the brane for
$\omega_\mu$ and $\curl\bar{\cal Q}_\mu$, on large scales. Using
Eqs.~(\ref{lc2'}), (\ref{pe4l}), (\ref{vec2}) and (\ref{vec}), we
find
 \bea
\dot{\bar{\alpha}}_\mu+\left(1-3c_{\rm s}^2\right)
H\bar{\alpha}_\mu&=&0\,,\\
\dot{\bar{\beta}}_\mu+(1-3w)H\bar{\beta}_\mu &=&
\left[{8\over3}H\left( 3c_{\rm s}^2-1\right) {{\cal U}\over\rho}-
\kappa^4(1+w)^2\rho^2\right]\bar{\alpha}_\mu\,,
 \eea
where
 \be
\bar{\alpha}_\mu=a\,\omega_\mu\,,~~
\bar{\beta}_\mu={a\over\rho}\curl{\cal Q}_\mu\,,
 \ee
are dimensionless vector perturbations. Thus we can solve for the
matter and nonlocal vorticity on large scales, without solving the
bulk perturbation equations, similar to the case of scalar
perturbations. The matter vorticity sources nonlocal vorticity,
but during slow-roll inflation this only leads to a highly damped
mode $\propto a^{-8}$. In very high-energy radiation domination
however, there is a weakly growing mode:
 \be
\bar{\alpha}_\mu=b_\mu\,,~~\bar{\beta}_\mu=c_\mu+Bb_\mu \ln
\left({a\over a_o}\right)\,,
 \ee
where $\dot B=\dot{b}_\mu=\dot{c}_\mu=0$.

\subsection{Tensor perturbations on the brane}

For the transverse traceless modes on the brane, the scalar
equations reduce to background equations, the vector equations
drop out and we can derive a covariant wave equation for the
tensor shear from the remaining
equations~(\ref{pe5l})--(\ref{pe7l}) and (\ref{pcc3l}):
\begin{eqnarray}
&& \D^2{\bar{\sigma}}_{\mu\nu}-\ddot{\bar{\sigma}}_{\mu\nu}
-5H\dot{\bar{\sigma}}_{\mu\nu}-\left[2\Lambda+{\textstyle{1\over2}}
\kappa^2(\rho-3p)-{\textstyle{1\over2}} \kappa^2
(\rho+3p){\rho\over\lambda}\right]{\bar{\sigma}}_{\mu\nu}\nonumber\\
&&~~{}=- {6\over\kappa^2\lambda}\left[ \dot{\bar{\cal P}}_{ \mu\nu
}+2H \bar{\cal P}_{\mu\nu} \right] \,. \label{t1}
\end{eqnarray}
In standard general relativity, the right hand side falls away,
and once $\bar{\sigma}_{\mu\nu}$ is determined, one can determine
$\bar{E}_{\mu\nu}$ from Eq.~(\ref{pe5l}) and $\bar{H}_{\mu\nu}$
from Eq.~(\ref{pcc3l}) (the shear is a gravito-electromagnetic
potential). Nonlocal bulk effects provide driving terms that are
like anisotropic stress terms in general relativity~\cite{c}. In
the latter case however, the evolution of anisotropic stress is
determined by the Boltzmann equation or other intrinsic physics.
The nonlocal anisotropic stress $\bar{\cal P}_{\mu\nu}$ from the
bulk Weyl tensor is not determined by brane equations, and can in
principle introduce significant changes to the shear via
Eq.~(\ref{t1}).

\section{Conclusion}

By adopting a covariant approach based on physical and geometrical
quantities that are in principle measurable by brane-observers, we
have given a classical analysis of intrinsic cosmological and
astrophysical dynamics in generalized Randall-Sundrum-type
brane-worlds. We have emphasised the role of the Weyl tensors in
the bulk and on the brane, and we have carefully delineated what
can and can not be predicted by brane observers without additional
information from the unobservable bulk. Intriguing new effects are
introduced by the extra dimension, and there are exciting problems
to be solved, especially on brane-world black holes and
gravitational collapse, and on cosmological perturbations and the
CMB and LSS.

\[ \]
{\bf Acknowledgments}

I would like to thank the organisers of EREs2000 for support and
warm hospitality, and David Wands and Filippo Vernizzi for helpful
comments.

\end{document}